 \documentclass[final,1p,times]{elsarticle}

\usepackage{graphicx}
\usepackage{xcolor}

\usepackage{amssymb}

\journal{Journal of LaTeX preprints}

\begin{document}

\begin{frontmatter}



\title{The PRad Windowless Gas Flow Target}


\author[JLab]{J. Pierce \fnref{ORNL}}
\author[JLab]{J. Brock}
\author[JLab]{C. Carlin}

\author[JLab]{C. Keith\corref{cor}}
\ead{ckeith@jlab.org}

\author[JLab]{J. Maxwell}
\author[JLab]{D. Meekins}
\author[UVa]{X. Bai}
\author[JLab]{A. Deur}
\author[MSU]{D. Dutta}
\author[Duke,TUNL]{H. Gao}
\author[NCAT]{A. Gasparian}
\author[UVa]{K. Gnanvo}
\author[Duke,TUNL]{C. Gu}
\author[JLab]{D. Higinbotham}
\author[NSU]{M. Khandaker}
\author[UVa]{N. Liyanage}
\author[Duke,TUNL]{M. Meziane}
\author[JLab]{E. Pasyuk}
\author[Duke,TUNL]{C. Peng \fnref{ANL}}
\author[NSU]{V.~Punjabi}
\author[Duke,TUNL]{W. Xiong \fnref{Syracuse}}
\author[Duke,TUNL]{X. Yan}
\author[MSU]{L. Ye}
\author[Duke,TUNL]{Y. Zhang}
\address[JLab]{Thomas Jefferson National Accelerator Facility, Newport News, VA 23606}
\address[MSU]{Mississippi State University, Mississippi State, MS 39762}
\address[Duke]{Duke University, Durham, NC 27708}
\address[TUNL]{Triangle Universities Laboratory, Durham, NC 27708}
\address[UVa]{University of Virginia, Charlottesville, VA 22904}
\address[UNCW]{University of North Carolina Wilmington, NC 28403}
\address[NCAT]{North Carolina A\&T State University, Greensboro, NC 27411}
\address[NSU]{Norfolk State University, Norfolk, VA 23504}
\cortext[cor]{Corresponding author}
\fntext[ORNL]{Present Address: Oak Ridge National Laboratory, Oak Ridge, TN 37830}
\fntext[Syracuse]{Present Address: Syracuse University, Syracuse, NY 13244}
\fntext[ANL]{Present Address: Argonne National Laboratory, Lemont, IL 60439}

\begin{abstract}
We report on a windowless, high-density, gas flow target at
Jefferson Lab that was used to measure $r_p$, the root-mean-square
charge radius of the proton. To our knowledge, this is the first such 
system used in a fixed-target experiment at a (non-storage ring) 
electron accelerator. The target achieved its design goal of an 
areal density of 2$\times$10$^{18}$ atoms/cm$^2$, with the gas uniformly 
distributed over the 4~cm length of the cell and less than 1\% 
residual gas outside the cell.  This design eliminated 
scattering from the end caps of the target cell, a problem endemic to
previous measurements of the proton charge radius in electron scattering experiments, and
permitted a precise, model-independent extraction of $r_p$ by reaching
unprecedentedly low values of $Q^2$, the square of the electron's transfer of four-momentum to the
proton.
\end{abstract}


\end{frontmatter}


\section{Introduction}
\label{sec:intro}
The Proton Radius Experiment at Jefferson Lab (PRad) \cite{Xiong2019} carried out 
a precise measurement of an important quantity in physics, the root-mean-square (rms) charge radius of the proton, $r_p$.  Precise knowledge of $r_p$ has a wide-ranging impact: from our understanding of the structure of the proton in terms of its quark and gluon degrees of freedom, to our knowledge of the 
Rydberg constant -- a fundamental constant of nature -- due to the impact $r_p$ has on 
bound-state quantum electrodynamics (QED) calculations of atomic energy levels.  The charge radius of
the proton can be measured using two techniques.  In the first,
it is extracted from spectroscopic measurements of energy level differences 
of the hydrogen atom (e.g.\ the Lamb shift), combined with state-of-the-art 
quantum electrodynamics (QED) calculations.  
In the second method, utilized by PRad, $r_p$ is determined
from the slope of the proton's electric form factor $G_E$, extracted from the electron-proton 
{\em e-p} elastic scattering cross section and extrapolated to zero momentum transfer.
More formally, $r_p$ is given by
\begin{equation}
    r_p =  \left( -6{\frac{d{G_{E}}}{d{Q^2}}}|_{Q^2=0} \right)^{1/2}
    \label{eqn:rp_define}
\end{equation}
where $Q^2$ is the square of the four-momentum transfer in {\em e-p} elastic scattering. 

Historically, $r_p$ obtained from these two methods agreed within experimental uncertainties~\cite{Mohr:2008}.  However, in 2010 $r_p$ was obtained for the first 
time from a measurement of the Lamb shift of muonic hydrogen, in which the
electron of the H atom is replaced by the much heavier muon. 
The result was a factor of ten more precise than all previous measurements~\cite{Pohl:2010}, 
but significantly {\em smaller} than previous measurements.  
Around the same time, a new electron scattering experiment was also performed with over 1400 data points at Mainz~\cite{Bernauer:2010} and a new value of $r_p$ was 
extracted.   Although the new result was more precise than previous scattering measurements, 
it was consistent with the old results, leading to a $>$7$\sigma$ discrepancy between the 
muonic hydrogen and regular hydrogen values of $r_p$.
This triggered the ``proton charge radius puzzle'' and led to major experimental and theoretical efforts to understand and/or resolve the discrepancy.
In this regard, significant progress has been made in recent years.
The latest Lamb shift results on regular hydrogen~\cite{Bezginov19} favor the
smaller value of $r_p$ indicated by muonic hydrogen.  Likewise,
the PRad result~\cite{Xiong2019} also agrees with the muonic hydrogen $r_p$.

The PRad experiment featured a number of innovations that made it the least model-dependent 
of all modern, high-precision electron scattering measurements of $r_p$ to date. 
First, utilizing a large-acceptance, high-resolution electromagnetic calorimeter (HyCal), it achieved the  
lowest $Q^2$ ever observed for {\em e-p} scattering in a magnetic-spectrometer-free measurement.  
Additionally, the large acceptance of the calorimeter allowed coverage in $Q^2$ that was
wide enough to ensure the necessary extrapolation to $Q^2=0$ 
in Eq.~\ref{eqn:rp_define} was robust.  The second innovation was the
simultaneous detection of {\em e-e} (M{\o}ller scattering) and {\em e-p} elastic
scattering in the same experimental acceptance. Doing so helped control systematic uncertainties
associated with the beam-target luminosity to an unprecedented level.  The third innovation, and the topic of this article, was a new hydrogen gas target that eliminated scattering from the end caps of the target cell, a problem common to previous electron scattering measurements of $r_p$. 
Together, these innovative methods permitted a precise electron scattering measurement 
at unprecedentedly low values of $Q^2$, allowing for extraction of $r_p$ in a model-independent manner.
The PRad result agrees with the muonic hydrogen results and gives
support to the recently revised value for the Rydberg constant \cite{Mohr2018}, 
one of the most accurately determined fundamental constants in nature.

Here we report on the design, construction, and performance of the windowless, cryo-cooled, continuous-flow hydrogen gas target that was used in the PRad experiment. The target incorporated a novel design feature of small apertures on the front and back surfaces of the target cell, such that the electron beam interacted almost exclusively with the hydrogen gas inside the cell. 
Gas that escaped through the apertures and into the accelerator beam line was 
removed by a number of high capacity vacuum pumps, reducing its density by three or more orders
of magnitude.
With this design, the target maintained an areal density of approximately
$2\times 10^{18}$ hydrogen atoms/cm$^2$ distributed uniformly over the 4~cm length of the target,
while also minimizing the amount of material exposed to the beam outside the cell, a critical factor for
reducing systematic uncertainties in the experiment.

\section{Target Design and Construction}  \label{Sec:Design}
\begin{figure*}
\begin{center}
\includegraphics[width=5in]{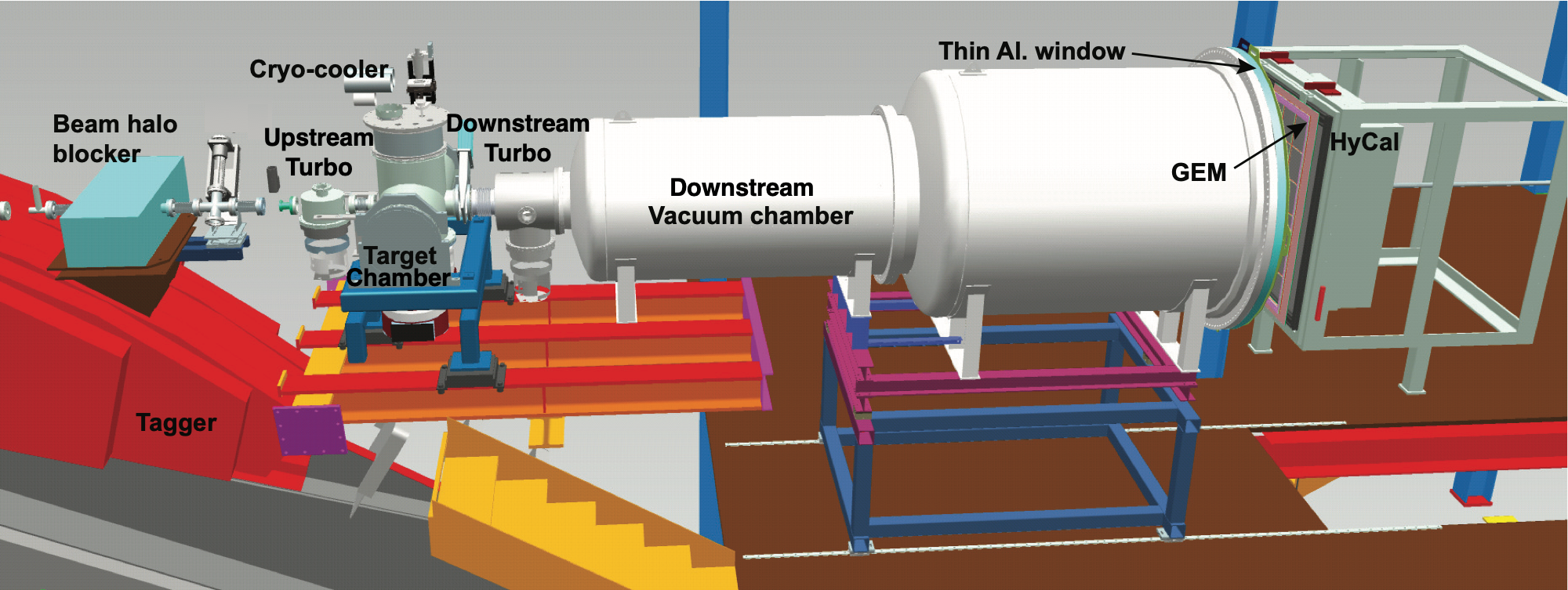}
\end{center}
\caption{Layout of the PRad experiment in Hall B at Jefferson Lab.  In this model, the electron
beam travels from left to right.}
\label{fig:PRadSetup}
\end{figure*}

To detect the energy and scattering angle of electrons in both
{\em e-p} and {\em e-e} scattering at very low values of Q$^2$, the PRad experiment 
(Fig.~\ref{fig:PRadSetup}) utilized HyCal, an electromagnetic hybrid calorimeter 
originally built for a precise measurement of the neutral pion radiative 
decay width by the PrimEx collaboration~\cite{HyCal,HyCal-PRL,HyCal-Science}.
The angular resolution of the measurements was further improved using 
two gas electron multiplier (GEM) detectors directly in front of HyCal.
Nevertheless, as the detectors were placed at very forward angles, it was not possible to 
reconstruct the scattering vertex with extreme precision. 
Furthermore, backgrounds are often a serious issue for very forward-angle
electron scattering experiments because the cross section for many 
processes increases with decreasing scattering angle.
These aspects made it critical to
localize the hydrogen target sample to a relatively small volume free from
any contaminants, including beam-entrance and beam-exit windows. 
At the same time, a highly accurate determination of the absolute target density 
was not necessary, thanks to the simultaneous measurement
of {\em e-e} rates from M{\o}ller scattering alongside the elastic {\em e-p} rates.
To this end, the PRad target was a sample of hydrogen gas flowing continuously through an
open (windowless) target cell 4~cm long.  The gas was cooled to cryogenic
temperatures to increase its volumetric density inside the cell to about
$5\times10^{17}$~atoms per~cm$^{3}$, and the cell was
specifically designed to create a large pressure difference between gas inside the cell 
and the surrounding beam line vacuum.

\begin{figure}
\begin{center}
\includegraphics[width=3.in]{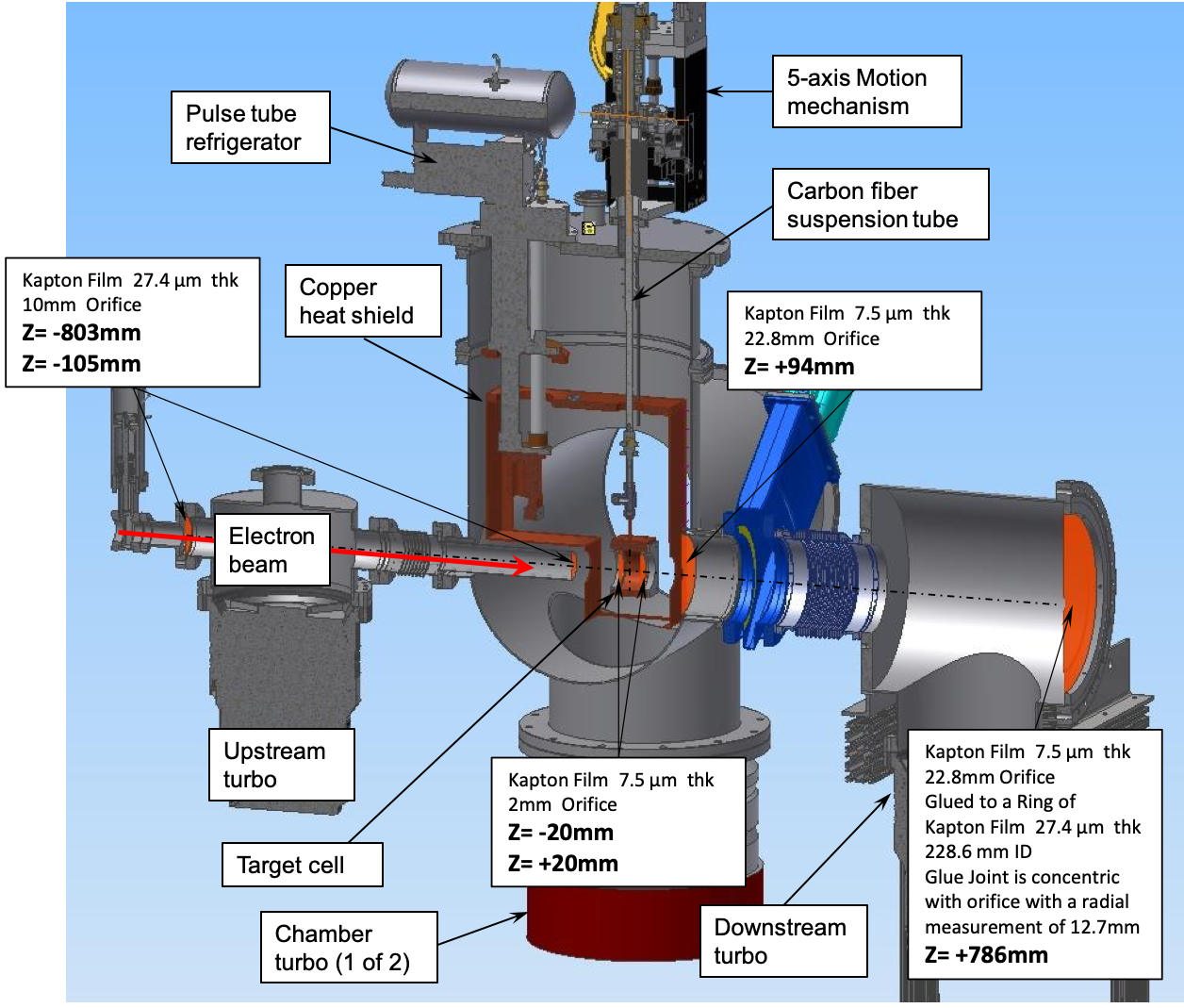}
\end{center}
\caption{Annotated drawing of the PRad gas flow target indicating most of the target's main components.
The location and dimensions of various polyimide (Kapton\textsuperscript\textregistered) pumping orifices are shown, where {\bf Z} is the distance
from target center.  The direction of the electron beam is indicated by a red arrow.}
\label{fig:Chamber}
\end{figure}

\begin{figure}
\begin{center}
\includegraphics[height=2in]{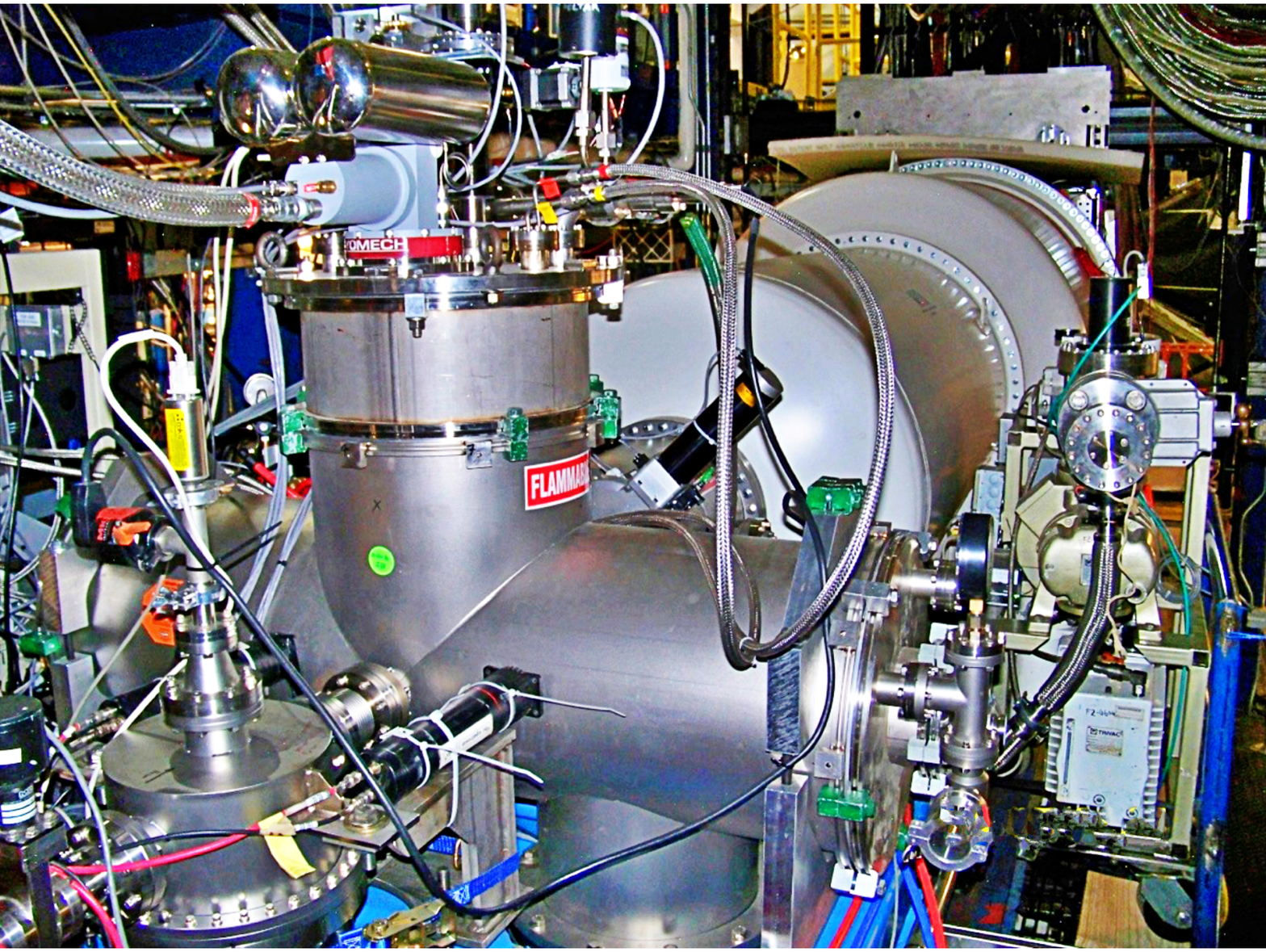}
\hspace{0.5in}
\includegraphics[height=2in]{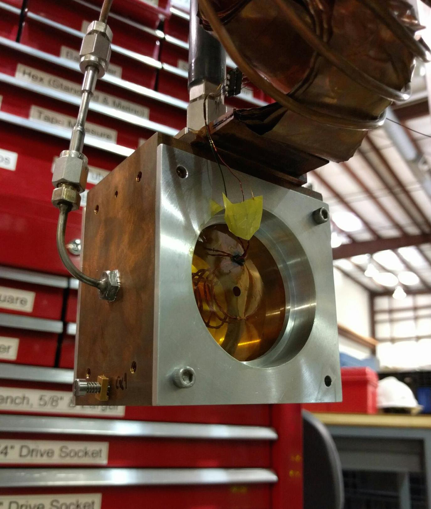}
\end{center}
\caption{Left: The Prad target system on the beam in Hall B.  In this view, the elecrton beam transverses the target from the lower left hand corner of the photo to the upper right hand corner.
Right: The PRad target cell.  Hydrogen gas, cooled by the pulse tube cryocooler, enters the cell via the
tube on the left.  The cell is cooled by a copper strap attached at the top, and is suspended by the carbon
tube directly above the cell.  A 2~mm orifice is visible at the center of the polyimide window, as are
wires for a thermometer inside the cell.  Two 1~$\mu$m solid foils of aluminum and carbon attach to the cell bottom but are not shown in the photograph. }
\label{fig:Photos}
\end{figure}

Figure~\ref{fig:Chamber} is a sectional drawing of the PRad target chamber and shows most of its
major components.  A photograph of the target installed on the Hall B beam line is shown 
on the left in Fig.~\ref{fig:Photos}.  High-purity hydrogen gas (\textgreater 99.99\%) was
supplied from a high-pressure cylinder located outside the experimental hall and
metered into the target system via a 0--10~slpm mass flow controller. Using a pair of
remotely actuated valves, the gas could be directed into the target 
cell for production data-taking, or into the top of the chamber for background measurements.  

Before entering the cell, the gas was cooled to cryogenic temperatures using a two-stage pulse tube
cryocooler (Cryomech model PT810) with a base temperature of 8~K and a
cooling power of 20~W at 14 K.  The cryocooler's first stage served two purposes.  
It cooled a tubular, copper heat exchanger that lowered the hydrogen gas temperature
to approximately 60~K.  It also cooled a copper heat shield surrounding 
the lower temperature components of the target, including the target cell.
The second stage of the cryocooler cooled the gas to its final operating temperature
using a similar heat exchanger and cooled the target cell via a 40~cm long, flexible copper strap.  
The temperature of the second stage was measured by a calibrated 
cernox thermometer\footnote{Lake Shore Cryotronics} and
stabilized at approximately 15~K using a small cartridge heater and
automated temperature controller. Without this,
the hydrogen gas would condense or even freeze inside the second stage heat exchanger.

The target cell, shown on the right in Figure~\ref{fig:Photos}, was machined from a single block of C101 copper.
Its outer dimensions were $7.5 \times 7.5 \times 4.0$~cm$^3$, with a 6.3~cm diameter hole along the
axis of the beam line.
The hole was covered at both ends by 7.5~$\mu$m thick polyimide foils held in place by aluminum
end caps.  Cold hydrogen gas flowed into the cell at its midpoint and exited via 2~mm holes
at the center of either polyimide foil.  The holes also allowed the electron beam to pass through the H$_2$
gas without interacting with the foils themselves, effectively making this a ``windowless'' gas target.
Compared to a long thin tube, the design of a relatively large target cell with small orifices 
had two important advantages.
First, it produced a more uniform density profile along the beam path,
allowing us to better estimate the gas density based upon its temperature and pressure. 
Second, it greatly reduced a potential source of background scattering.
Rather than scattering from the 4~cm long copper cell walls, any ``halo'' electrons outside the primary beam radius could only scatter from the much thinner 
7.5~$\mu$m polyimide foils.

A second calibrated cernox thermometer, suspended inside the cell, provided a direct measure of
the gas temperature. Approximately 50~cm of each of the thermometer's four lead wires
was coiled inside the cell to improve the thermal conduction between the thermometer and gas. 
The gas pressure was measured by a 0--10~torr capacitance manometer located
outside the vacuum chamber and
connected to the cell by a carbon fiber tube approximately one meter long and 2.5~cm in diameter.
The same tube was used to suspend the target cell from a motorized, 5-axis motion controller 
which could position the target with a precision of about $\pm10$~$\mu$m.  
The motion controller was also used to lift the cell out of the beam
in order to investigate possible scattering of beam halo from the polyimide windows.  Finally,
two 1~$\mu$m thick carbon and aluminum foils were attached to the bottom of the 
copper target cell for background and calibration measurements. 

High-speed turbomolecular pumps were used to evacuate the hydrogen gas
as it left the target cell and maintain the surrounding vacuum chamber 
and beam line at low pressure.  Two Pfeiffer HiPace 3400 magnetically
levitated turbo pumps, each with a nominal pumping speed of
3000~l/s, were attached directly under the chamber, 
while two additional Pfeiffer HiPace 1500 turbo pumps with 1400~l/s speed each were used on the upstream and downstream portions of the beam line.  A second capacitance manometer measured the hydrogen gas pressure inside the target chamber, while cold cathode vacuum gauges were utilized in all other locations.  
While the response of capacitance manometers was independent of the gas species being measured, the cathode gauge readings required correction for the ionization energy of hydrogen, made according to the manufacturer's specifications.

As illustrated in Fig.~\ref{fig:Chamber}, additional polyimide orifices were installed at various locations to limit the extent of target (hydrogen) gas along the path of the beam.  With this design, the density of gas decreased significantly outside the target cell, 
with an estimated 99\% of scattering occurring within the 4~cm length of the cell (Sec~\ref{sec:Production}). 
For obvious reasons of safety, the hydrogen exhausted from all vacuum pumps was vented outside the experimental hall. A continuous flow of nitrogen gas was also added to the vent line to prevent the formation of a combustible mixture of hydrogen and oxygen.

\section{Target Performance} \label{sec:Performance}
\begin{figure*}
\begin{center}
  \includegraphics[width=.4\linewidth]{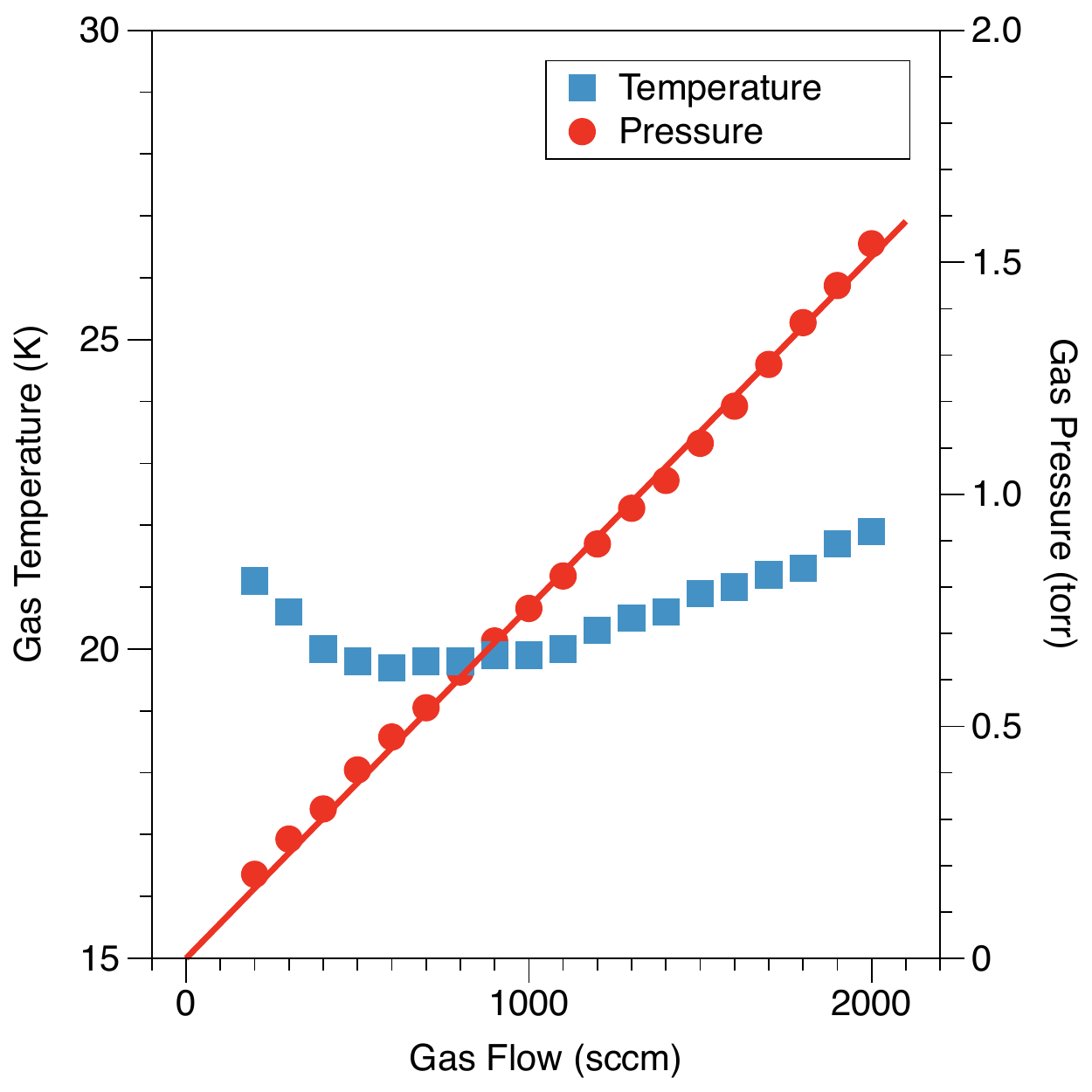} \hspace{0.25in}
  \includegraphics[width=.4\linewidth]{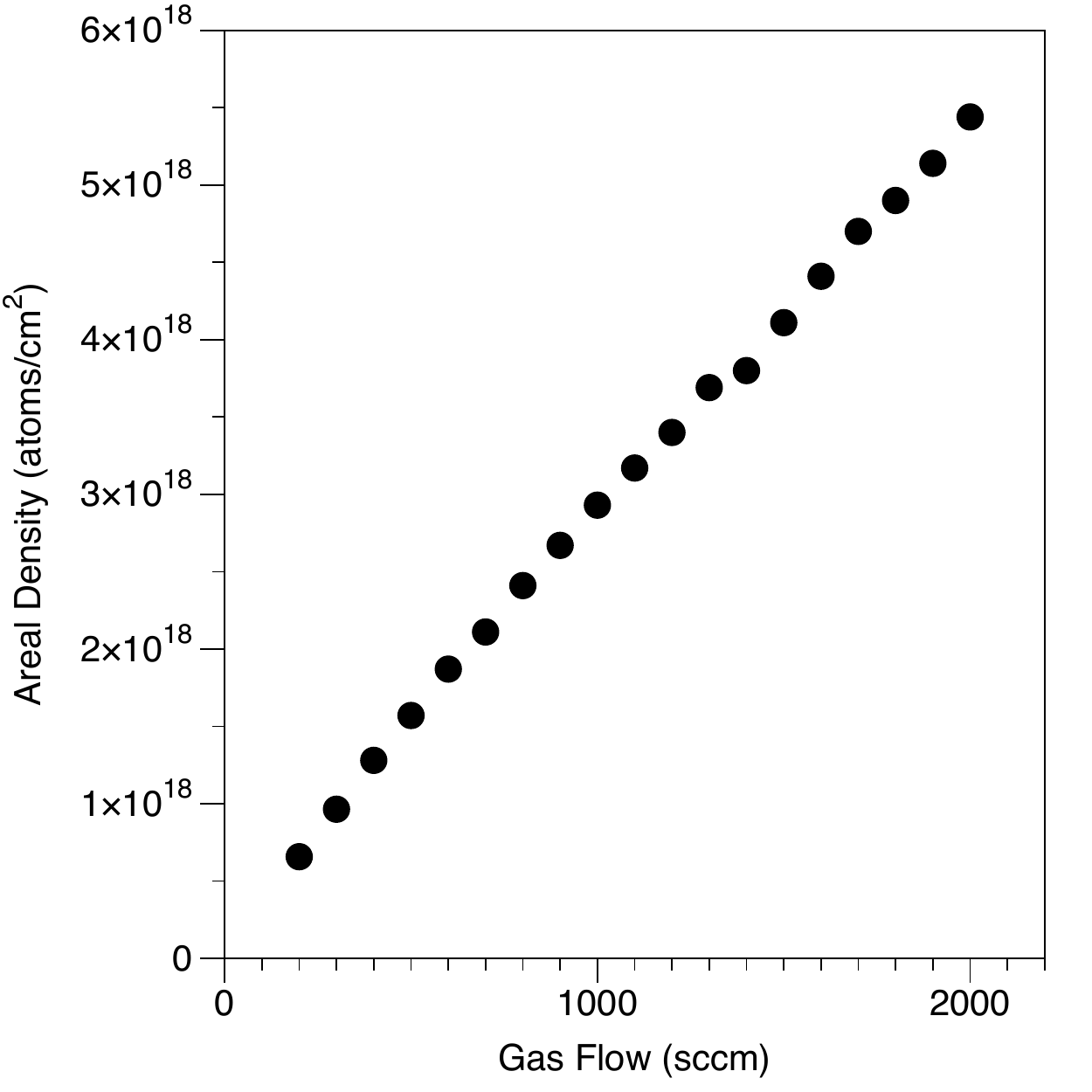}
  \end{center}
\caption{Left: Temperature (blue squares) and pressure (red circles) of hydrogen gas within the PRad target
cell as a function of gas flow through the cell.  The red line is a calculation of the gas flow
using Eq.~\ref{eqn:ChokedFlow} and a constant discharge coefficient $C=0.65$.  More details in the text.
Right: Corresponding areal density as a function of gas flow based on the measured pressure and temperature. }
\label{fig:GasFlow}
\end{figure*}

The temperature and pressure of H$_2$ gas flowing through the PRad target cell, as well as the resulting areal density, are shown as a function of flow rate in Fig.~\ref{fig:GasFlow}.
For these measurements, the temperature of the cryocooler was regulated at 15~K.
At lower temperatures, target operation became unstable as hydrogen condensed and eventually froze inside the second stage heat exchanger.   The cernox thermometer inside the cell had a calibration
accuracy of $\pm 9$~mK, while the accuracy of the capacitance manometer was 
$\pm0.01$~torr.  No attempt was made to determine temperature gradients within the cell.
The pressure difference between cold gas in the cell and room temperature gas in the manometer
was estimated using the correlation function~\cite{Thermomolecular}:
\begin{equation}
	\frac{P_H - P_L}{P_L} = 2\times10^{-9} \left(r P_L\right)^{-1.99} \left[T^{2.27}_H - T^{2.27}_L\right]
\label{eqn:Thermomolecular}
\end{equation}
and was less than 0.2\% under all measured conditions.  
Here $P_{H,L}$ and $T_{H,L}$ are the
pressures and temperatures of the gas at the High and Low temperature ends of the connecting tube of radius $r$, expressed in Pa, K, and m, respectively.

As shown in Fig.~\ref{fig:GasFlow}, the temperature of the gas inside the cell
was largely independent of flow rate, while the pressure increased in a linear manner. 
This is the expected behavior of a compressible, near-ideal gas flowing
through an orifice of diameter $d_2$, where the mass flow rate
can be written as  \cite{OrificePlate}
\begin{equation}
	\dot m = \frac{\epsilon \, C}{\sqrt{1-\beta^4}} \; \frac{\pi}{4}\,d_2  \;\sqrt{2 \rho_1 (P_1 - P_2)}.
\label{eqn:ChokedFlow}
\end{equation}
Here $\rho_1$ is the density of the gas on the upstream side of the orifice, 
and $P_1$ and $P_2$ are its pressures on the upstream and
downstream sides, respectively.  $C$ is the
discharge coefficient (about 0.6 for an orifice with sharp edges),
$\beta = d_2/d_1$ is the ratio of the orifice diameter to the upstream pipe diameter, and $\epsilon$ is the 
expansibility factor for small-bore orifices
\cite{Buckingham}, 
\begin{equation}
	\epsilon = 1 - \frac{P_1 - P_2}{\gamma P_1} \left( 0.41 + 0.35 \beta^4 \right)
	\label{eqn:epsilon}
\end{equation}
with  $\gamma$ the ratio of the gas's specific heats.  For
hydrogen gas at the PRad operating conditions, $\gamma=C_p/C_v=1.66$.
Taking $P_1 \gg P_2$, $\beta \ll 1$, and $\rho_1 \propto P_1$, Eq.~\ref{eqn:ChokedFlow} 
reduces to the linear relationship between pressure and flow that is seen in Fig.~\ref{fig:GasFlow}.  
The red curve in Fig.~\ref{fig:GasFlow} was generated using Eq.~\ref{eqn:ChokedFlow} to 
calculate the flow of H$_2$ gas through two 2~mm orifices at the measured pressures and 
temperatures and using a discharge coefficient $C=0.65$.

\section{Target Operation} \label{sec:Operation}
\begin{figure}
\begin{center}
\includegraphics[width=4in]{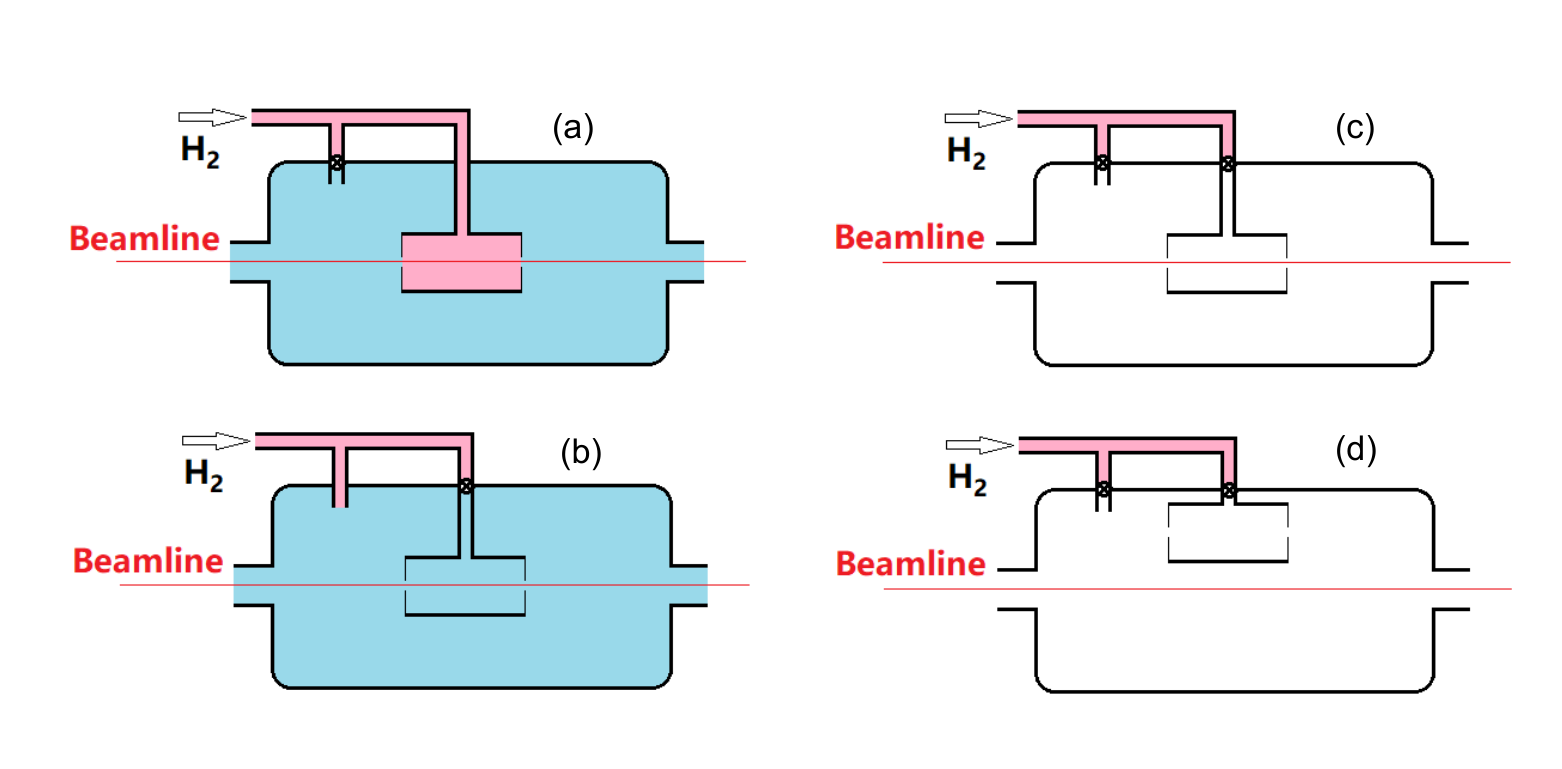}
\end{center}
\caption{Four configurations of the PRad target.  In each, pink indicates a region of high density hydrogen gas, blue indicates low-density hydrogen, and white indicates high vacuum. }
\label{fig:FourTargetModes}
\end{figure}

Data collection during the PRad experiment was typically
broken into one hour segments, or ``runs'', 
with the target operating in one of the four configurations 
illustrated in Fig.~\ref{fig:FourTargetModes}.  
Production data for measuring the proton charge radius
utilized configuration (a), in which 
high-density H$_2$ gas flowed through
the target cell while the surrounding vacuum chamber 
and beam line were filled with lower-density gas escaping from the cell.  
The performance the target in this configuration is described in Sec.~\ref{sec:Production}.
Configurations (b)--(d) were utilized to examine scattering of electrons
from material other than hydrogen atoms in the target cell and are the subject of
Sec.~\ref{sec:Background}.

\subsection{Production Run Performance} \label{sec:Production}
\begin{figure*}
\begin{center}
  \includegraphics[width=.35\linewidth]{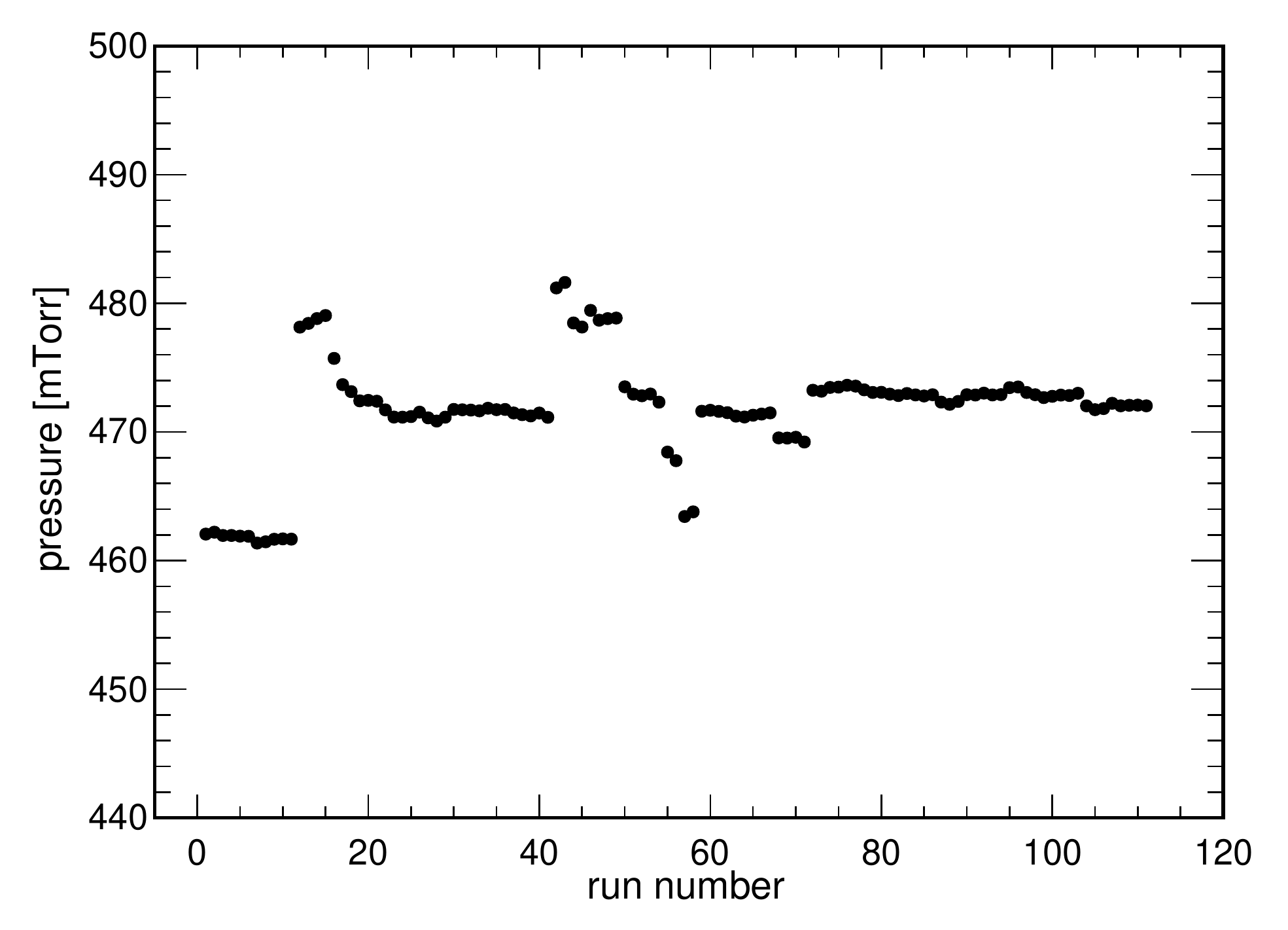} \hspace{0.25in}  \includegraphics[width=.35\linewidth]{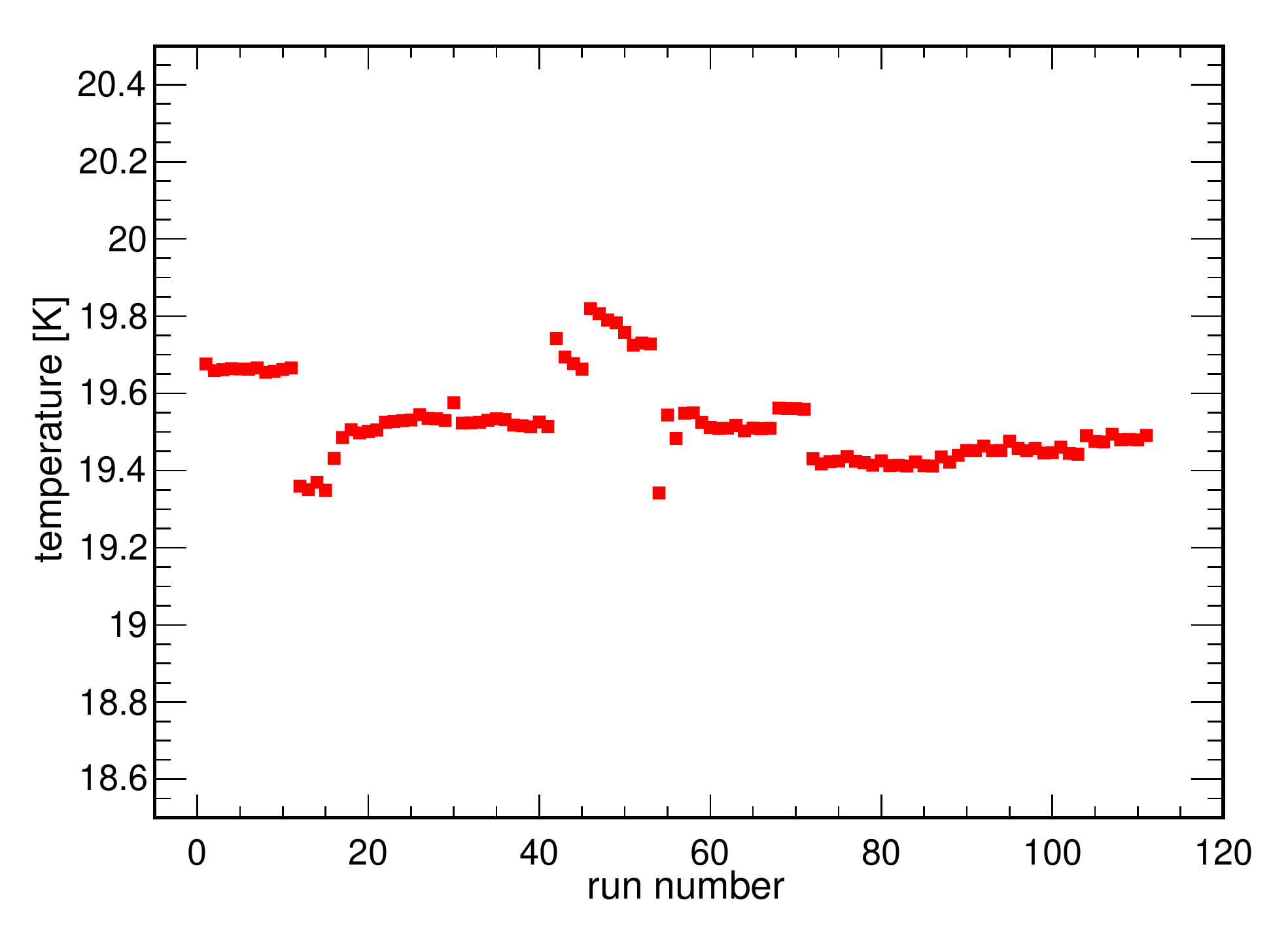}
\includegraphics[width=.35\linewidth]{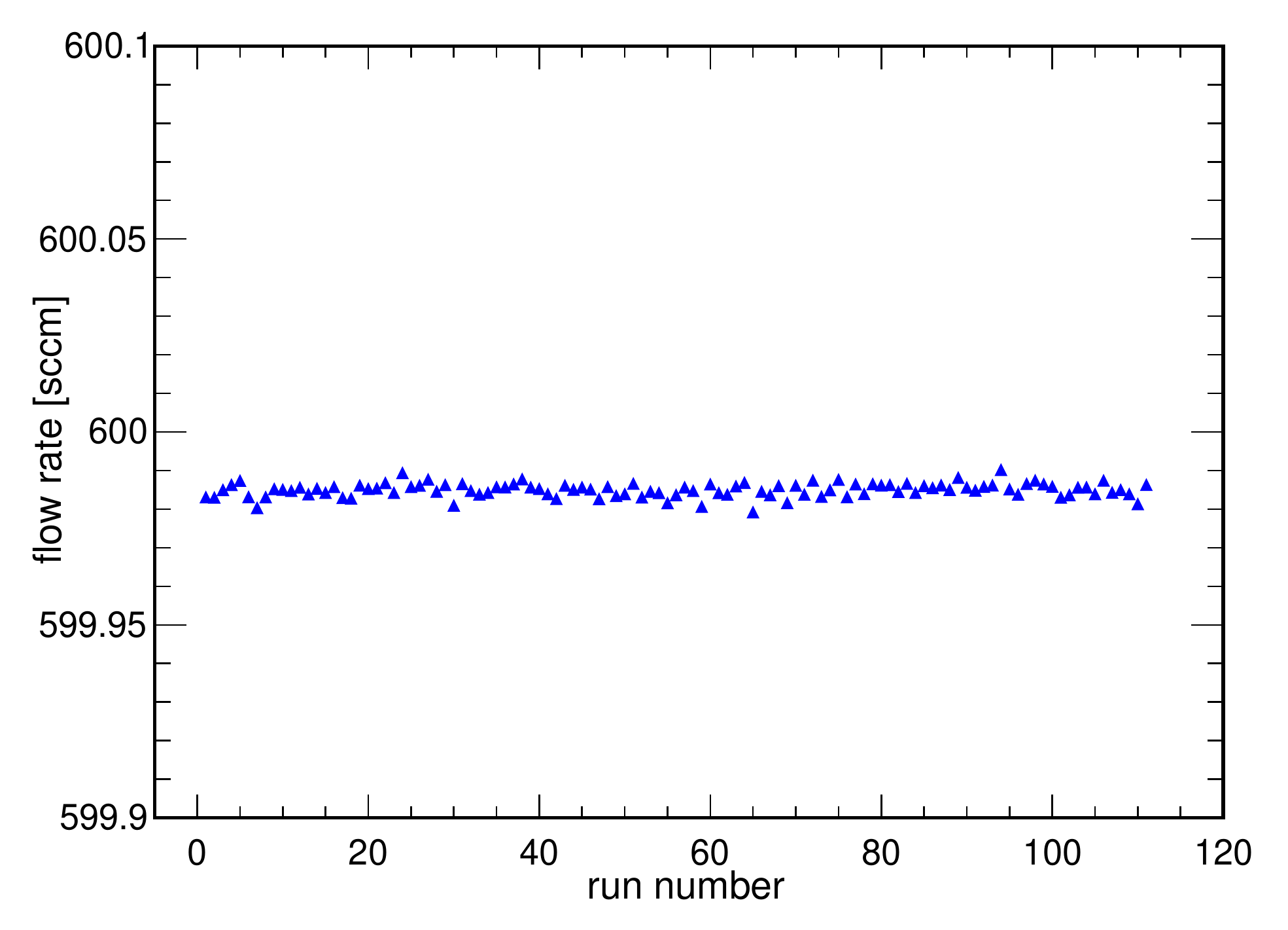} \hspace{0.25in}
  \includegraphics[width=.35\linewidth]{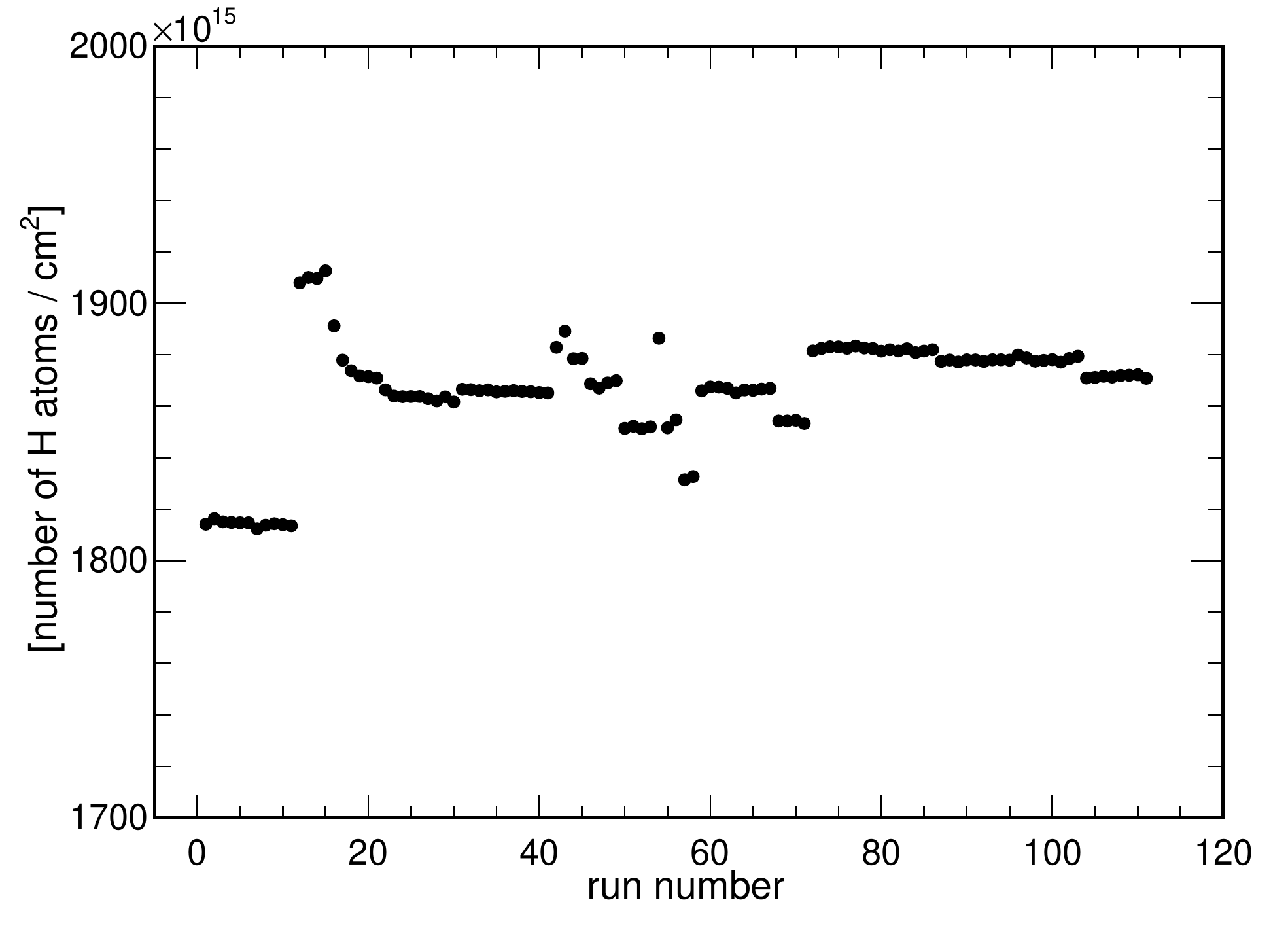}
  \end{center}
\caption{Performance of the PRad target during all full target
(configuration a) runs. Clockwise from upper left: cell gas pressure,
cell gas temperature, areal density of gas in target cell, and gas flow into the cell.}
\label{fig:Stability}
\end{figure*}

All production runs for measuring $r_p$ 
were made with 600 sccm H$_2$ gas flowing through 
the target cell, giving pressure and temperature measurements 
of about 0.47~torr and 19.5~K, respectively.
The resulting gas density was 0.78~$\mu$g/cm$^3$ \cite{NIST},  
which corresponded to a $1.9 \times 10^{18}$~cm$^{-2}$
areal density of hydrogen atoms within the 4~cm long cell. 
The performance of the target throughout all 110 production runs is shown in Fig.~\ref{fig:Stability}.
During the course of any one hour run, the gas temperature
and pressures varied by less than one percent,
although fluctuations up to a few percent between runs
can be seen in Fig.~\ref{fig:Stability}.  These occurred following long periods
of operation with other target configurations but had no impact on the extracted value of $r_p$
because the {\em e-p} elastic scattering rates were always normalized to the M{\o}ller scattering rates.

Gas pressures measured in other regions of the beam line
(``residual gas'') were two-to-four
orders of magnitude lower than the cell pressure (Table~\ref{table:pressures}).
The greatest quantity of residual gas along the beam
path was inside the 4~m long downstream vacuum chamber (Fig.~\ref{fig:PRadSetup}). 
Here the pressure was slightly higher than at the downstream turbo pump, presumably due to
outgassing or leaks in the chamber.
This can be greatly reduced in future installations
by additional pumping on the chamber or
reducing the 22.9~mm orifice at the 
chamber's entrance (see Fig.\ref{fig:Chamber}).  

\begin{figure}
\begin{center}
  \includegraphics[width=.8\linewidth]{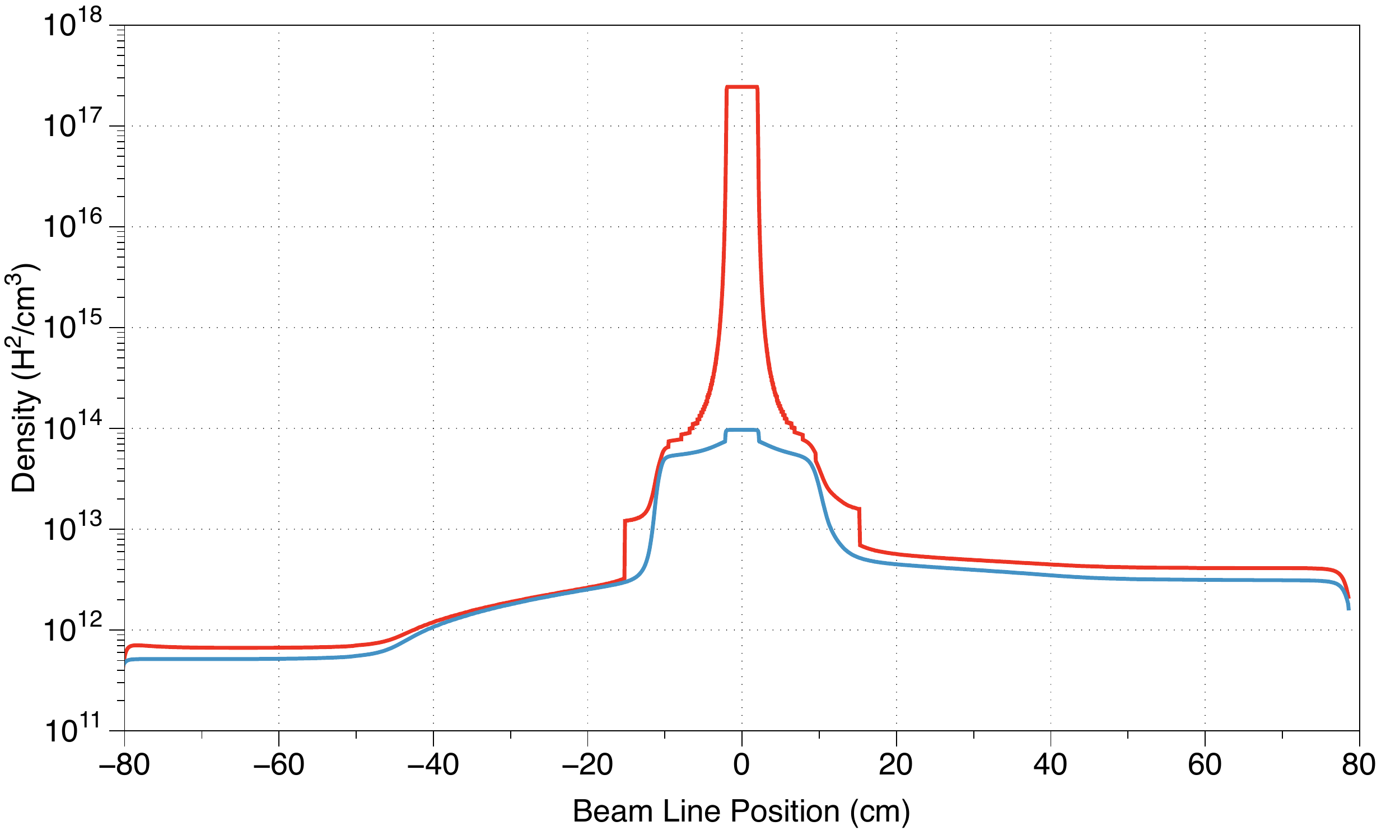}
  \end{center}
\caption{COMSOL simulation of the hydrogen density along the beam line.  Red: H$_2$ gas is cooled
and injected into the target cell at nominal PRad operational settings.  Blue: H$_2$ gas is injected directly into the vacuum chamber surrounding the target cell.  The cell center is located at 0~cm, with positive values downstream of the target.}
\label{fig:DensityPlot}
\end{figure}

Table~\ref{table:pressures} indicates that approximately 99\% of all hydrogen 
in the beam's path was constrained within the 4~cm length
of the target cell.  Because the pressure sensors were mounted several centimeters from
the beam axis, the values in Table~\ref{table:pressures} could not be utilized
to accurately correct for the presence of the residual gas.  Instead,
these corrections were made using the background measurements described below.
In addition, the COMSOL Multiphysics\textsuperscript{\textregistered} 
modelling software was used to simulate
 the density of H$_2$ gas flowing through the target system and beam line in configurations
 (a) and (b) (Figure~\ref{fig:DensityPlot}).  
 Additional studies, including simulations with various density profiles outside the
 target cell, were performed, and systematic uncertainties were assigned to account
 for the presence of the residual gas~\cite{Xiong-thesis}.  This, along with 
 halo scattering contributed a systematic uncertainty of less
 than 0.5\%~\cite{Nature_Supplement} to the extracted value of $r_p$.

\begin{table}[htp]
\scriptsize
\begin{center}
\begin{tabular}{|c||c|c|c|c|}
\hline
Beam Line	   	& Length	& Pressure	& Thickness	 	& Percentage \\
Region 		& (cm)  	& (torr)  	& (atoms/cm$^2$)  	& of total \\
\hline \hline
Target Cell	&  4		& 0.47				&  $1.9 \times 10^{18}$     & 99.06 \\
\hline
US Beam line	&  300 	& $2.2 \times 10^{-5}$	& $4.4 \times 10^{14}$	& 0.02 \\
\hline
US Turbo		& 71		&  $5.7 \times 10^{-5}$	& $2.7 \times 10^{14}$	& 0.01 \\
\hline
Target Chamber  & 14	&  $2.3 \times 10^{-3}$	& $2.1 \times 10^{15}$	& 0.11 \\
\hline
DS Turbo     	& 71		&  $3.0 \times 10^{-4}$	& $1.4 \times 10^{15}$	& 0.07 \\
\hline
DS Chamber 	& 400	& $5.2 \times 10^{-4}$	& $1.4 \times 10^{16}$	& 0.72 \\
\hline
\end{tabular}
\end{center}
\label{table:pressures}
\caption{Hydrogen gas pressures and thickness (areal densities) for the PRad beam  at the nominal gas flow rate of 600~sccm.  US and DS refer to Up- and Down-stream portions of the beam line, relative to the target cell.
See Fig.~\ref{fig:PRadSetup} for more details.  Room temperature gas is assumed in calculating the areal density of all regions
except Region 1 (target cell), where a temperature of 19.5~K is used.}
\end{table}

\subsection{Background Measurements} \label{sec:Background}
The target configurations (b), (c), and (d) shown in Fig.~\ref{fig:FourTargetModes} were used
to study sources of background in the PRad measurements, that is, electrons that scattered from
material other than hydrogen atoms inside the target cell.
In configuration (b), the hydrogen gas flow was kept at 600 sccm 
but was admitted directly into the target chamber
rather than the target cell.   Thus, all scattering sources along the beam path 
were the same as in production runs except for gas inside the cell, 
which was reduced more than three orders of magnitude.
The resulting charge-normalized data rates for {\em e-p} and {\em e-e} (M{\o}ller) scattering 
made with background configuration (b) were then subtracted from the full-cell measurements to isolate
scattering from hydrogen atoms within the target cell.  
Configurations (c) and (d) were used to better
understand the origin of background events.
There was no gas flowing into the system in either configuration, and the only
difference was the location of the target cell.  The cell remained in the beam path in (c) but was lifted
in (d), thus removing the cell windows as a possible source of background.

\begin{figure*}
\begin{center}
  \includegraphics[width=.6\linewidth]{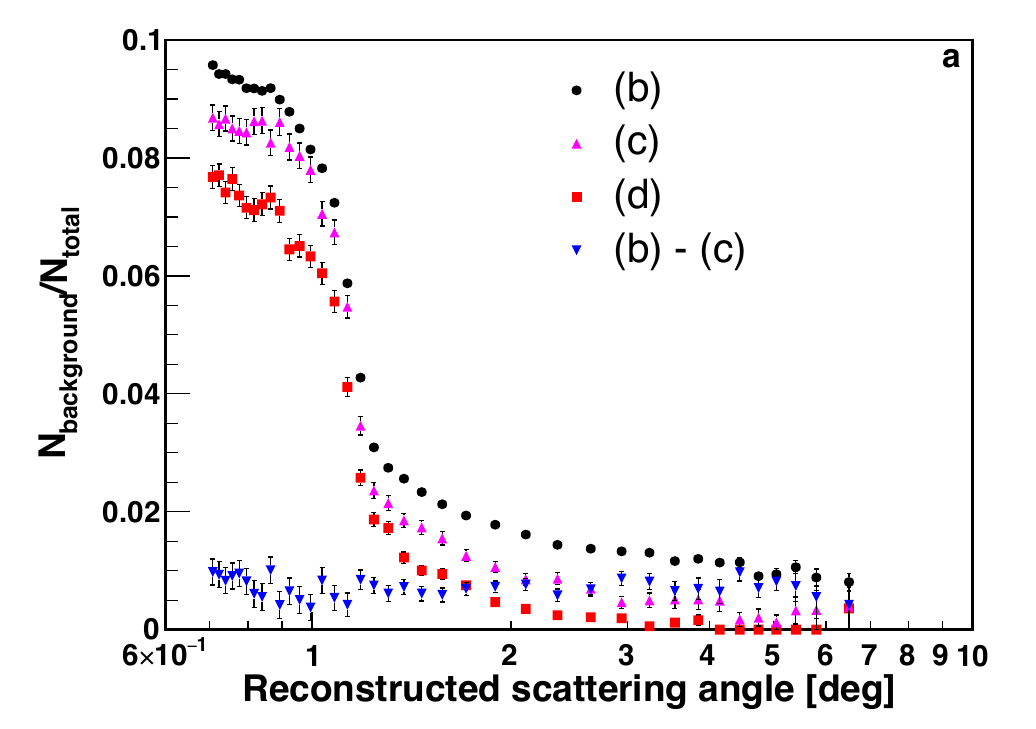}
  \end{center}
\caption{e-p scattering rates during background measurements as a function of reconstructed 
electron scattering angle at a beam energy of 2.2~GeV.  
The target configurations 
(b, c, d) are indicated in the plot legend.  In all cases, the rates are
normalized to the scattering rates of the full target configuration (a).}
\label{fig:ScatteringPlots}
\end{figure*}

Scattering rates for each of the three background configurations 
are plotted as a function of reconstructed electron scattering angle in Fig.~\ref{fig:ScatteringPlots}.
These measurements were made at a 2.2~GeV beam energy and normalized
to the production scattering rates measured with configuration (a).
All rates display prominent peaks at very forward angles, indicating the greatest
sources of background scattering were near or upstream from the target cell.
As expected, the rates from configuration (b) were the greatest, since they included
all sources of background scattering, including residual hydrogen gas in the beam line.
The background contribution from this residual gas
can be determined from the difference (b)-(c) and is seen
to be approximately 1\%, consistent with the results shown in Table~\ref{table:pressures}.

Rates for configurations (c) and (d) are similar, which indicates little background
from the target cell windows.  We conclude that the majority of the background
(6--8\%) came from halo scattering
from beam line elements other than the target and was likely
produced by the upsteam Beam Halo Blocker seen in Fig.~\ref{fig:PRadSetup},
a 12.7-mm diameter collimator designed to reduce the intrinsic size of the halo.

\section{Summary}  \label{Sec:Summary}
We have described a new hydrogen gas target utilized in PRad,
an electron scattering measurement of the root-mean-squared
charge radius of the proton conducted
at Jefferson Lab.  The target design eliminated the beam entrance and exit windows
that have constituted major sources of background scattering in previous
$r_p$ measurements from electron scattering.  
Together with other innovative instrumentation and measurement
techniques, the target permitted a precise and model-independent 
extraction of $r_p$ from {\em e-p\/} elastic scattering.  
This target will be used in a newly approved PRad-II~\cite{PRad-II} experiment at JLab that will improve the proton charge radius measurement by a factor of nearly four compared with the PRad experiment.
The apparatus described here is also compatible with practically any noncorrosive target gas (deuterium,
helium, argon, neon, etc.\/), and can be used in other experiments where such a target system is advantageous.

\section*{Acknowledgements}
This work was funded in part by the US National Science Foundation (NSF MRI PHY-1229153) and by the US Department of Energy (contract number DE-FG02-03ER41231), including contract number DE-AC05-06OR23177, under which Jefferson Science Associates, LLC operates the Thomas Jefferson National Accelerator Facility. We thank the staff of Jefferson Laboratory,
in particular the Jefferson Lab Target Group, for their support throughout the experiment. We are also grateful to all grant agencies for providing funding support to the authors throughout this project. 





\bibliographystyle{elsarticle-num}
\bibliography{<your-bib-database>}



\end{document}